\pdfoutput=1
\documentclass{article}
\usepackage{cite}
\usepackage{amsmath,amssymb}
\usepackage{spconf,graphicx}
\usepackage{algorithm}
\usepackage{algorithmic}
\usepackage{hyperref}
\usepackage{makecell, multirow}
\usepackage{bbm}
\usepackage{booktabs}

\title{Partially Fake Audio Detection by Self-attention-based \\Fake Span discovery}

\name{\begin{tabular}{c}
    Haibin Wu$^1$,
    Heng-Cheng Kuo$^2$,
    Naijun Zheng$^4$,
    Kuo-Hsuan Hung$^2$ \\
	Hung-Yi Lee$^1$,
	Yu Tsao$^2$,
	Hsin-Min Wang$^3$,
	Helen Meng$^4$
\end{tabular}}

\address{
$^1$ Graduate Institute of Communication Engineering, National Taiwan University \\
$^2$ Research Center for Information Technology Innovation, Academia Sinica, Taiwan \\
$^3$ Institute of Information Science, Academia Sinica, Taiwan \\
$^4$ Human-Computer Communications Laboratory, The Chinese University of Hong Kong \\
}

\begin{document}
\ninept
\maketitle
\begin{abstract}
The past few years have witnessed the significant advances of speech synthesis and voice conversion technologies. However, such technologies can undermine the robustness of broadly implemented biometric identification models and can be harnessed by in-the-wild attackers for illegal uses.
The ASVspoof challenge mainly focuses on synthesized audios by advanced speech synthesis and voice conversion models, and replay attacks.
Recently, the first Audio Deep Synthesis Detection challenge (ADD 2022) extends the attack scenarios into more aspects.
Also ADD 2022 is the first challenge to propose the partially fake audio detection task.
Such brand new attacks are dangerous and how to tackle such attacks remains an open question.
Thus, we propose a novel framework by introducing the question-answering (fake span discovery) strategy with the self-attention mechanism to detect partially fake audios.
The proposed fake span detection module tasks the anti-spoofing model to predict the start and end positions of the fake clip within the partially fake audio, address the model's attention into discovering the fake spans rather than other shortcuts with less generalization, and finally equips the model with the discrimination capacity between real and partially fake audios.
Our submission ranked second in the partially fake audio detection track of ADD 2022.
\end{abstract}

\section{Introduction}
\label{sec:intro}

The past few years have witnessed significant advances in speech synthesis and voice conversion technologies, and recently emerged adversarial attacks, such that even humans may not be capable to distinguish the real users’ speech from the synthesised speech \cite{wu2015asvspoof,kinnunen2017asvspoof,todisco2019asvspoof, yamagishi2021asvspoof, Yi2022ADD, wu2020defense_2, wu2020defense, peng2021pairing, wu2021adversarial, liu2019adversarial, li2021replay, li2021channel, wu2021voting, wu2021improving, wu2021spotting, wu2015spoofing,wu2014study,kamble2020advances,das2020assessing, chenglong2021global,ma2021continual,yi2021half,wang2021comparative}.
Such technologies can undermine the robustness of broadly implemented biometric identification models, e.g. automatic speaker verification (ASV) models, and can be harnessed by in-the-wild attackers for criminal usage.
For instance, an attacker can generate fake audios to manipulate the voiceprint-based security entrance system to accept the attacker falsely, and get access to normally protected information and valuables.
Additionally, an imposter can call the bank center, fool the biometric identification system to accept him/her as a registered user, and transfer money to the imposter's account. 
Considering the severe harm caused by synthesized fake audio, it is critical to devise methods to tackle such threats. 

The ASVspoof challenge \cite{wu2015asvspoof,kinnunen2017asvspoof,todisco2019asvspoof, yamagishi2021asvspoof}, a community-led challenge, arouses the attention from both the industry and the academia to tackle the spoofing audios in both physical access and logical access.
In logical access, attacks are mainly from synthesized audios by advanced speech synthesis and voice conversion models, while in physical access, replayed audios are adopted as attacks.
The challenge attracts various international teams, and various high-performance anti-spoofing models have been proposed to address the two kinds of attacks mentioned above.
The adversarial attacks for ASV and anti-spoofing models have been well investigated \cite{wu2020defense_2, wu2020defense, peng2021pairing, wu2021adversarial,wu2021improving, wu2021spotting}.
To solve further challenging attack situations in realistic applications, the first Audio Deep Synthesis Detection challenge (ADD 2022) \cite{Yi2022ADD} extends the attack scenarios to fake audio detection.
They consider the fake audios perturbed by diverse background noise, and attacks from the latest speech synthesis and voice conversion models.
Additionally, the organizers propose partially fake audio detection track, where the attacks are composed of hiding small fake clips into real speech.
Partially fake audios are dangerous, and ADD 2022 is the first challenge attempting to tackle this type of brand new attacks, which is an open question, and is the focus of this paper

\begin{figure*}[ht]
  \centering
  \centerline{\includegraphics[width=0.9\linewidth]{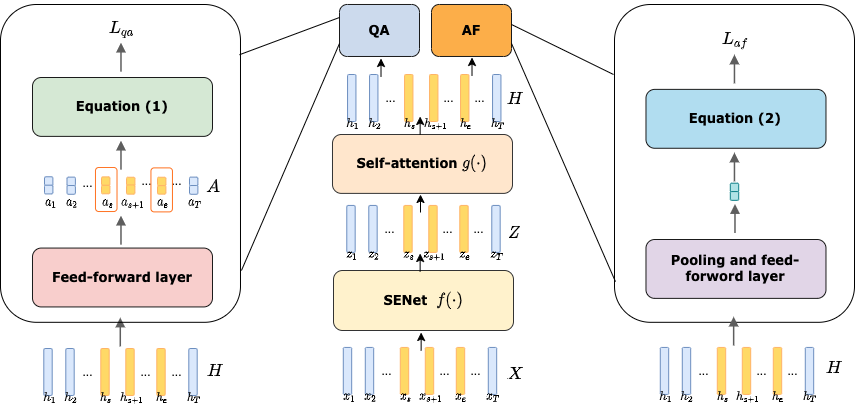}}
  \vspace{0.2cm}
  \caption{The proposed framework. $X,Z,H,A$ are the acoustic features, hidden features, bottleneck features and the output for Question-answer layer, respectively. $f$ and $g$ are the SENet feature extractor and self-attention layer, corresponding to (1)-(8) and (9) in Table~\ref{tab:model}, respectively. 
QA and AF are the question-answering (fake span discovery) and anti-spoofing layers with loss calculation procedures respectively.}
  \label{fig:framework}
\end{figure*}

During generation of partially fake audio, only small clips of synthetic speech are inserted, and thus the fake audio contains a large proportion of genuine user's audio.
Through experimentation, we find it is hard to distinguish the fake and real audios by directly implementing the previous state-of-the-art spoofing countermeasure models, such as Light Convolutional Neural Network (LCNN) \cite{lavrentyeva2017audio} and Squeeze-and-Excitation Network (SENet) \cite{lai2019assert}.
To allow the model discover the small anomalous clip in real speech, we design a proxy task to make the model answer ``where are the start and end points'' of such anomalous clips.
During training, the anti-spoofing model not only has to predict the fake or real label for each utterance, but also to find the start and end positions of the fake clips within the utterance. 
Identifying the time segments of the fake clips is similar to \textit{extraction-based question-answering}, which determines the answer span in a document.
Also, to further improve the capacity of the anti-spoofing model to tackle the ``question-answering'' task, we introduce the self-attention \cite{vaswani2017attention} strategy.
The experimental results illustrate the effective discrimination capacity of our proposed method between real and partially fake audios.

Our main contributions are two-fold:
\begin{itemize}
    \item We proposed a brand new framework inspired by the extraction question-answering strategy for locating the fake regions in the fake, overall input audio, in order to improve the performance for partially fake audio detection.
    \newpage
    \item We further equipped the fake span discovery strategy with the self-attention mechanism to get a better detection capacity.
\end{itemize}
Also, our submission ranked second in the partially fake audio detection track of ADD 2022

This paper is organized as follows:
Section 2 introduces the proposed method, namely self-attention-based question-answering framework for partially fake audio detection 
Section 3 presents experimental setups, followed by section 4 reporting on experimental results and analysis.
Section 5 presents the conclusion.

\section{Methodology}
\label{sec:method}
In this section, we will introduce the anti-spoofing method equipped with the proposed question-answering strategy and self-attention mechanism.
We firstly present the details of the proposed framework.
And then we will clarify the rationale of the proposed method.

\subsection{Proposed anti-spoofing model}

We adopt the base model SENet \cite{lai2019assert}, which is a variant of ResNet \cite{he2016deep} equipped with squeeze-and-excitation networks \cite{hu2018squeeze}, and we perform some modifications to that model.
The modified model architecture is shown in Table~\ref{tab:model}.
Let $X=[x_{1},x_{2},...x_{T}]$ denote the $T$ frames of input acoustic features.
The extracted hidden features by the SENet feature extractor are denoted as $f(X)=Z = [z_{1}, z_{2}, ...z_{T}]$, where $f$ is the (1)-(8) layers in Table~\ref{tab:model}.
The bottleneck features are denoted as $g(Z)= H = [h_{1}, h_{2}, ...h_{T}]$, where $g$ is the self-attention layer, the layer (9) in Table~\ref{tab:model}, and $h_{t} \in R^{n}$.
The self-attention layer is one layer of transformer \cite{vaswani2017attention}.
The question-answering layer (a) is one fully-connected layer with the input dimension as $n$, the dimension for $h_{t}$, and with output dimension as 2.
The 2 dimensions represent how likely will $h_{t}$ be the start or end position of the fake clip.
Given $H$ as the input, the question-answering layer will output $A = [a_{1}, a_{2}, ...a_{T}]$, where $a_{t} \in R^{2}$.
The question-answering loss $L_{qa}$ is denoted as:
\begin{equation}
    L_{qa}=-(log \frac{exp(a_{s}^{1})}{\sum_{t=1}^{T} exp(a_{t}^{1})} + log \frac{exp(a_{e}^{2})}{\sum_{t=1}^{T} exp(a_{t}^{2})}),
    \label{eq:qa-loass}
\end{equation}
where $s$ and $e$ are the start and end positions for the fake clip, $a_{t}^{1}$ and $a_{t}^{2}$ are the values for first and second dimensions of $a_{t}$ at the $t^{th}$ frame.
We will not incorporate the $L_{qa}$ for training with real utterances.
For the pooling layer (b), there are three pooling strategies in this paper, average pooling, self-attentive pooling (SAP) \cite{bhattacharya2017deep} and attentive statistics pooling (ASP) \cite{okabe2018attentive}.
Based on the bottleneck features $H$, the pooling layer (b) followed by the prediction layer (c) will output $S=[s_{0}, s_{1}]$, indicating whether the utterance is fake or real.
The anti-spoofing loss $L_{af}$ is denoted as:
\begin{equation}
    L_{af}= -log \frac{exp(s_{l})}{\sum_{j=0}^{1} exp(s_{j})},
    \label{eq:af-loass}
\end{equation}
where $l \in \{0,1\}$ is the target label.
The final loss is 
\begin{equation}
    L=L_{qa}+L_{af}.
    \label{eq:final-loss}
\end{equation}

\begin{table}[t!]
    \centering
    \small
    \caption{Proposed anti-spoofing model.}
    \vspace{0.3cm}
    \begin{tabular}{cc|c|l}
    
    \toprule
     layer & Type &  Filter / Stride & Output shape \\
    \cmidrule(r){1-4}
    (1) & Conv      & $7\times7 / 1\times2$ & $16\times501\times40$ \\
    (2) & BatchNorm & $-$ & $-$                     \\
    (3) & ReLU      & $-$ & $-$                    \\
    (4) & MaxPool   & $3\times3 / 1\times2$ & $16\times501\times20$ \\
    \hline
    (5) & SEResNet Module$\times3$ & $-$ & $16\times501\times20$    \\
    \hline
    (6) & SEResNet Module$\times4$ & $-$ & $32\times501\times10$    \\
    \hline
    (7) & SEResNet Module$\times6$ & $-$ & $64\times501\times5$     \\
    \hline
    (8) & SEResNet Module$\times3$ & $-$ & $128\times501\times3$    \\
    \hline
    (9) & Self-attention           & $-$ & $501\times384$ \\
    \toprule
    (a) & Question-answering       & $-$ & $501\times2$   \\
    \hline
    (b) & Pooling                  & $-$ & $384$    \\
    \hline
    (c) & Prediction               & $-$ & $2$      \\
    \bottomrule
    \end{tabular}
    \label{tab:model}
\end{table}

\subsection{Rationale}
\label{subsec:model}
In the partially fake audio detection track, there is only a small proportion of fake audio frames in the overall piece of input speech. 
Previous state-of-the-art anti-spoofing models \cite{lai2019assert,lavrentyeva2017audio} tackle the problem of identifying whether a whole audio utterance is real or fake. 
Hence, previous strategies are not designed to identify anomalous regions within one utterance.
Thus the previous models intuitively attain the ability to distinguish between utterances but there is no guarantee that such models can discover the abnormal regions within a single utterance. 
To evaluate the performance of the previous state-of-the-art anti-spoofing models, we direct train binary classification anti-spoofing models for the partially fake audio detection task with reference to previous papers.
We discover that these well-trained models do not have the discriminative ability for the adaptation set provided by the organizers of ADD 2022.
A reasonable explanation is that the models may have learned some shortcuts to differentiate the audios with real and fake labels in the training set, but what the models have learned can not generalize to the adaptation set.
In other words, the models cannot discover the fake regions for fake audio detection.

Thus, to regularize the model to learn to distinguish between the real and partially fake audios, we propose a proxy task to let the model discover the abnormal parts within a piece of partially fake audio.
The proposed anti-spoofing model has to predict not only whether the input utterance is real or fake, but also output the start and end of each anomalous region.
We name this proxy task as question-answering, or fake span discovery proxy task, in which the model has to answer ``where is the fake clip'' in a piece of partially fake audio.
The extraction-based question-answering models in natural language processing (NLP) often take a question and a passage as input, build representations for the passage and the question respectively, match the question and passage embeddings, and output the start and end positions within the passage as the answer.
We adopt the analogy of extraction-based question-answering here. 
The passage is the partially fake utterance, and the answer span is the time of the fake clip.
By the question-answering proxy task, the model can learn to find the fake clips within an utterance, thus benefiting the model to distinguish between the audios with or without fake clips.
Moreover, the self-attention module followed by the question-answering task addresses the model to attend on the fake regions, and helps reduce the question-answering loss, resulting in better discrimination capacity between real and partially fake audios.

\section{Experimental setup}
\label{sec:setup}

\subsection{Data preparation}
\subsubsection{Dataset construction}


The training set and dev set, which are based on Mandarin publicly available corpus AISHELL-3 \cite{shi2020aishell}, provided by the organizers of ADD 2022, cannot be directly adopted to tackle the problem of partially fake audio detection track, as the whole utterance sample in them is either real or fake.
During the training phase, for constructing fake audios, we generate the partially fake audio by inserting a clip of audio into the real audios.
The inserted clips are derived from three sources: 1). fake audios in the training and dev set provided by ADD 2022. 2). Real audios other than the victim audio in the training and dev set. 3). audios re-synthesised by the traditional vocoders, including Griffin-Lim \cite{griffin1984signal} and WORLD \cite{morise2016world}, based on the real audios in the training and dev set.
It is hard to train text-to-speech (TTS) or voice conversion (VC) models based on the limited real data provided by the organizer, so we choose the traditional vocoders, namely Griffin-Lim and WORLD, to increase the diversity of fake audios.
As for the validation set, we adopt the adaptation set consisting of partially fake audios synthesised by ADD2022 for selecting the models.
We report the equal error rate (EER) for the testing set released by the organizer, as EER is the main evaluation metric for the partially fake audio detection track.

\begin{table}[htb]
    \caption{The EERs with (w/) or without (w/o) self-attention.}
    \vspace{0.3cm}
    \centering
    \begin{tabular}{cccc}
    \toprule
        FFT window size & w/o attention & w/ attention \\
        \cmidrule(r){1-3}
        384 & 23.6\% & 14.3\% \\
        768 & 22.0\% & 17.9\% \\
    \bottomrule
    \end{tabular}
    \label{tab:eer for self-attn}
\end{table}

\subsubsection{Input representations} 
Mel-spectrograms, which are based on short-time Fourier transform (STFT) where the window size of fast Fourier transform (FFT) is varied from 384 to 768, the hop size is 128, and the number of output bins is 80, are used as input features for most of our experiments and are denoted by MSTFT in following sections.
Besides spectral features, some extra experiments are operated on cepstral and NN-based features to increase diversity for achieving a better performance in the stage of fusion. 
The FFT window size, hop size, and number of output bins are fixed to 384, 128, and 80 respectively for Mel-frequency cepstral coefficients (MFCC), linear frequency cepstral coefficients (LFCC), and SincNet \cite{ravanelli2018speaker}, as we find the FFT window size of 384 performs well as shown in Table~\ref{tab:eer for mstft}.

\subsubsection{Data augmentation} 
We perform on-the-fly data augmentation by adding noise from MUSAN dataset \cite{snyder2015musan}, performing room impulse response (RIR) simulation \cite{ko2017study} and applying codec algorithms (a-law and $\mu$-law) \cite{recommendation1988pulse}. 

\subsection{Implementation details}
The backbone model is shown in Table~\ref{tab:model}.
Three kinds of attention, average pooling (Avg), attentive statistics pooling (ASP) and self-attentive pooling (SAP) are adopted for experiments.
All the models are optimized by Adam with the learning rate of 0.001 and weight decay as $1e^{-4}$.

\begin{table*}[htb]
    \caption{The EERs using MSTFT features. w/o or w/ mean with or without. w/ or w/o re-synthesis correspond to using the re-synthesised audios by Griffin-Lim and WORLD or not.}
    \vspace{0.5cm}
    \centering
    \begin{tabular}{ccccccc}
    \toprule
        \multirow{2}{*}{feature} & \multirow{2}{*}{FFT window size} & \multirow{2}{*}{pooling method} & \multicolumn{2}{c}{w/o augmentation} & \multicolumn{2}{c}{w/ augmentation} \\
        \cmidrule(r){4-5}\cmidrule(r){6-7}
        & & & w/o re-synthesis & w/ re-synthesis & w/o re-synthesis & w/ re-synthesis \\
        \cmidrule(r){1-3}\cmidrule(r){4-5}\cmidrule(r){6-7}
        \multirow{4}{*}{MSTFT} & 384 & Avg & 14.3\% & 19.9\% & 11.9\% & 14.2\% \\
                               & 512 & Avg & 13.2\% & 20.5\% & 13.0\% & 14.8\% \\
                               & 640 & Avg & 18.5\% & 19.9\% & 18.9\% & 13.3\% \\
                               & 768 & Avg & 17.9\% & 16.8\% & 14.8\% & 12.6\% \\
        \cmidrule(r){1-3}\cmidrule(r){4-5}\cmidrule(r){6-7}
        \multirow{4}{*}{MSTFT} & 384 & SAP & 16.9\% & 17.5\% & 15.6\% & 12.6\% \\
                               & 512 & SAP & 17.0\% & 18.0\% & 13.9\% & 12.5\% \\
                               & 640 & SAP & 12.1\% & 15.3\% & 15.3\% & 11.1\% \\
                               & 768 & SAP & 15.2\% & 17.8\% & 11.7\% & 14.8\% \\
        \cmidrule(r){1-3}\cmidrule(r){4-5}\cmidrule(r){6-7}
        \multirow{4}{*}{MSTFT} & 384 & ASP & 17.3\% & 15.9\% & 14.9\% & 11.9\% \\
                               & 512 & ASP & 14.9\% & 15.8\% & 12.9\% & 11.1\% \\
                               & 640 & ASP & 17.5\% & 15.9\% & 15.8\% & 11.2\% \\
                               & 768 & ASP & 14.8\% & 17.9\% & 14.5\% & 22.1\% \\
    \bottomrule
    \end{tabular}
    \label{tab:eer for mstft}
\end{table*}

\section{Experimental results and analysis}
\label{sec:expt}

\begin{table}[htb]
    \caption{The EERs for three different features}
    \vspace{0.3cm}
    \centering
    \begin{tabular}{cccc}
    \toprule
        feature &  MFCC & LFCC & SincNet  \\
        EER  & 12.5\% & 11.1\% & 16.1\% \\
    \bottomrule
    \end{tabular}
    \label{tab:eer for input}
\end{table}

First of all, we illustrate that the question-answering (QA) strategy drastically decreases the EERs.
We conduct experiments with and without the QA strategy. 
The experimental results show that the trained models without the QA strategy attain the EERs of around 40\%, which indicates that such models can not distinguish the partially fake audios from the genuine audios.
Due to the poor performance of models without the QA strategy on the adaptation set, we decide not to submit the results on testing sets of such models to the leaderboard to save the submission times.

Next, we verify the effectiveness of the self-attention layer by Table~\ref{tab:eer for self-attn}.
As the input and output feature dimensions after the self-attention layer are the same, the model without the self-attention layer can be constructed by directly removing (9) in Table \ref{tab:model}. 
In the following experiments, the performances on the testing set will be directly displayed.
We show the EERs under two settings of FFT window size due to space limitation, and the other settings are with the same trend. 
Table \ref{tab:eer for self-attn} shows that the improvements are significant in two settings with different window sizes. 
The EERs decrease 9.3\% and 4.1\% absolute after adding self-attention for the FFT window size of 384 and 768 respectively, which illustrates the significant improvements by introducing the self-attention layer.


Therefore, the model with self-attention will be adopted for the following experiments, unless specified otherwise.
In the main experiments as shown in Table~\ref{tab:eer for mstft}, the input representations are MSTFTs with hop size of 128, output bins as 80, and FFT window size ranging from 384-768.
Table \ref{tab:eer for mstft} exhausts the experimental settings under four different window sizes, three pooling strategies, whether to use the data augmentation and whether to use the re-synthesised fake audios by Griffin-Lim and WORLD.
We have the following observations.
First, EERs are improved with the help of data augmentation in most of the setups. 
Secondly, enlarging the training set by the re-synthesised data usually benefits the EERs when data augmentation is conducted. 
Lastly, the SAP and ASP pooling significantly improve the EERs when both data re-synthesis and augmentation are applied.
We also can observe that the best EER for a single model is 11.1\% shown in Table~\ref{tab:eer for mstft}.



In order to increase diversity of models for achieving a better performance in the stage of model fusion, we further take MFCC, LFCC and SincNet as input features to train the models. 
We cannot exhaust all the settings due to limited computing resources, thus we refer to Table~\ref{tab:eer for mstft} to select the setting to conduct the experiments.
We fix the FFT window size as 384, apply only ASP pooling, adopt data augmentation and the re-synthesised data. 
We observe from Table~\ref{tab:eer for input} that the LFCC feature gets EER as 11.1\%, reaching the best single model performance in our experimental settings.
For the further work, we plan to explore the potential of different front-end features to get better performance.

For the fusion method, we tried average fusion, weighted average fusion, min fusion and max fusion.
The best submission, which is fused by the average scores of the top 5 models, achieves the best 7.9\% EER and ranks second in partially fake audio detection track.

\section{conclusion}
\label{sec:conclusion}
Inspired by extraction-based question answering, this paper proposes a self-attention-based, fake span discovery strategy.
The proposed strategy tasks the anti-spoofing model to predict the start and end position of the fake clip within the partially fake audio, address the model's attention into discovering the fake spans rather than other patterns with less generalization, and finally equips the model with the discriminate capacity between real and partially fake audios.
Our final submitted model gave 7.9\% EER, and ranked 2nd in the partially fake audio detection track of ADD 2022.
Such a strategy can be model-agnostic and feature-agnostic.
Our future work will explore the potential of the proposed strategy by adopting other backbone anti-spoofing models and front-end features.

\section{acknowledgement}
\label{sec:acknowledge}
This research is partially funded by the Centre for Perceptual and Interactive Intelligence, an InnoCentre of The Chinese University of Hong Kong
This work was done while Haibin Wu was a visiting student at Human-computer Communications Laboratory, The Chinese University of Hong Kong.

\newpage


\bibliographystyle{IEEEbib}
\bibliography{strings,refs}

\end{document}